\documentclass{sigchi}

\CopyrightYear{2020}
\setcopyright{acmcopyright}
\doi{10.1145/3313831.3376321}
\isbn{978-1-4503-6708-0/20/04}
\conferenceinfo{CHI '20}{CHI Conference on Human Factors in Computing Systems, April 25--30, 2020, Honolulu, HI, USA}
\acmPrice{\$15.00}

\usepackage{tabularx}

\usepackage{balance}       
\usepackage{graphics}      
\usepackage[T1]{fontenc}   
\usepackage[utf8]{inputenc}
\usepackage{txfonts}
\usepackage{mathptmx}
\usepackage[pdflang={en-US}]{hyperref}
\usepackage{color}
\usepackage{booktabs}
\usepackage{textcomp}
\usepackage{subcaption}
\usepackage{tabu}
\usepackage[dvipsnames]{xcolor}
\usepackage{tikz}
\usepackage[subtle]{savetrees}
\usepackage{microtype}        
\usepackage{ccicons}          

\usepackage{todonotes}
\definecolor{diagramgreen}{RGB}{12,97,0}
    \def\plaintitle{Dark Patterns after the GDPR: Scraping Consent Pop-ups and Demonstrating their Influence}

\def\emptyauthor{}
\def\plainkeywords{Notice and Consent; Dark patterns; Consent Management Platforms; GDPR; Web scraper; Controlled experiment}

\makeatletter
\def\url@leostyle{%
  \@ifundefined{selectfont}{
    \def\UrlFont{\sf}
  }{
    \def\UrlFont{\small\bf\ttfamily}
  }}
\makeatother
\def\pprw{8.5in}
\def\pprh{11in}

\definecolor{linkColor}{RGB}{6,125,233}
\hypersetup{%
  pdftitle={\plaintitle},
  pdfauthor={\emptyauthor},
  pdfkeywords={\plainkeywords},
  pdfdisplaydoctitle=true, 
  bookmarksnumbered,
  pdfstartview={FitH},
  colorlinks,
  citecolor=black,
  filecolor=black,
  linkcolor=black,
  urlcolor=linkColor,
  breaklinks=true,
  hypertexnames=false
}

\begin{document}

\title{\plaintitle}

\numberofauthors{1}
\author{
 \alignauthor \ ~~~~~Midas Nouwens\textsuperscript{1,2}~~~Ilaria Liccardi\textsuperscript{2}~~~~Michael Veale\textsuperscript{3}~~~David Karger\textsuperscript{2}~~~~Lalana Kagal\textsuperscript{2}\\
 \email{\textnormal{}~~~~~~\{midasnouwens\}~~~~~~~~~~~~\{ilaria\}~~~~~~~~~~~~~~~~~\{m.veale\}~~~~~~~~~~~~~\{karger\}~~~~~~~~~~~~~\{lkagal\}~~~~~~~~}\\ \hspace{1cm}
  \begin{tabular}{cccccccccc}
     \affaddr{\textsuperscript{1}\{\}@cavi.au.dk{\ }} & & & &   \affaddr{\textsuperscript{2}\{\}csail.mit.edu {\ }}  & & & & & \affaddr{\textsuperscript{3}\{\}ucl.ac.uk{\ }}\\
    \affaddr{Digital Design \& Information Studies{\ }}  & & & &  \affaddr{~~~~MIT CSAIL {\ }}  & & & & & \affaddr{Faculty of Laws {\ }}\\
    \affaddr{Aarhus University, DK}  & & & & \affaddr{~Cambridge, MA, USA}  & & & & &\affaddr{UCL, UK {\ }} \\
  \end{tabular}
}


\maketitle

\begin{abstract}

New consent management platforms (CMPs) have been introduced to the web to conform with the EU's General Data Protection Regulation, particularly its requirements for consent when companies collect and process users' personal data.
This work analyses how the most prevalent CMP designs affect people's consent choices. 
We scraped the designs of the five most popular CMPs on the top 10,000 websites in the UK (n=680). 
We found that dark patterns and implied consent are ubiquitous; only 11.8\% meet the minimal requirements that we set based on European law. 
Second, we conducted a field experiment with 40 participants to investigate how the eight most common designs affect consent choices.
We found that notification style (banner or barrier) has no effect; removing the opt-out button from the first page increases consent by 22--23 percentage points; and providing more granular controls on the first page decreases consent by 8--20 percentage points. 
This study provides an empirical basis for the necessary regulatory action to enforce the GDPR, in particular the possibility of focusing on the centralised, third-party CMP services as an effective way to increase compliance.

\end{abstract}





\section{Introduction}
The predominant method of giving people some semblance of control over their privacy while browsing the web is `notice and choice' or `notice and consent'~\cite{cranorNecessaryNotSufficient2012} . 
These mechanisms involve showing an individual an informational statement and, depending on their (in)action, acquiring or assuming their agreement to collecting, storing, and processing their data. 
To many, this practice has become informally known as `cookie banners'.

What counts as sufficient notice, and what counts as legally-acceptable consent, significantly differs depending on the geographical and regulatory scope that an actor falls in. 
The application in Europe of the General Data Protection Regulation (GDPR)~\cite{gdpr} from May 2018, together with recent regulatory guidance from data protection authorities (DPAs) and jurisprudence from the Court of Justice of the European Union (CJEU), has highlighted the illegality of the way `notice and consent' has hitherto functioned in the EU. 
These regulatory changes have both clarified the concept of consent in European law, as well as brought more significant (and extraterritorial) consequences for flaunting these rules. 
EU law in particular focuses on the \emph{quality} of the consent required, and its freely-given, optional nature.

\emph{Consent management platforms} (CMPs) have gained traction on the Web to help website owners outsource regulatory compliance. These (often third-party) code libraries purport to help websites establish a lawful basis to both read and write information to users' browsers and to process these individuals' personal data, often for the purposes of tracking and complex advertising transactions, such as `real-time bidding'~\cite{informationcommissionersofficeUpdateReportAdtech2019}.

This intertwining of interface designs and data protection and privacy law raises significant questions. This paper deals with two of them:
\begin{enumerate}
    \item What is the current state of interface design of CMPs in the EU, and how prevalent are non-compliant design elements?
    \item How do interface designs affect consent actions of users and, by extension, how `freely given' that consent is?
\end{enumerate}

To answer the first question, we surveyed the designs of the 5 most commonly used third-party CMPs by scraping their varied implementations on the top 10,000 most popular websites in the United Kingdom (UK) (n=680); and evaluated them against European law and regulatory guidance.
To answer the second question, we built a browser plugin that injects consent notices into webpages and ran a controlled experiment (n=40) with eight different interfaces to see how they affect participants' consent responses. 

\section{Consent and Web Technologies under EU Law}

EU law considers users' devices and information within them part of their private sphere. Relevant protection is extended to all EU residents and to all individuals around the world being delivered online services from the Union. In light of a growing trend in the early 2000s of rightsholders sneaking piracy-spotting rootkits onto users' devices~\cite{kostaPeekingCookieJar2013a}, the ePrivacy Directive~\cite{eprivadirective} was amended to require that storing or accessing information on a user's device not `strictly necessary' for providing an explicitly requested service requires both clear and comprehensive information and opt-in consent~\cite{kostaPeekingCookieJar2013a}. This also applies to cookies, HTML web storage, and fingerprinting in browsers providing non-essential features such as tracking. Such consent is however, \emph{not} required for essential functions such as remembering login status, a shopping cart, or cookies for data security required by law~\cite{cnilcookies,informationcommissionersofficeGuidanceUseCookies2019}. 

The ePrivacy Directive is connected to definitions in European data protection law, so when the GDPR~\cite{gdpr} repealed and replaced the Data Protection Directive 1995~\cite{dpd} in 2018, these practices became subject to new, heightened standards concerning the quality of consent. The GDPR defines consent as ``any freely given, specific, informed and unambiguous indication of the data subject's wishes by which he or she, by a statement or by a clear affirmative action, signifies agreement to the processing of personal data relating to him or her''~\cite[art 4(11)]{gdpr}. Several aspects of this legal regime with design implications are important to highlight here, which are drawn from the legal texts, regulators' guidance, and court cases or opinions.

\subsection{Freely given and unambiguous consent}

Regulators and the Court of Justice of the EU have both emphasised that for consent to be freely given and informed, it must be a separate action from the activity the user is pursuing~\cite{planet49ag,article29workingpartyGuidelinesConsentRegulation2018,planet49court}. So-called `implicit' or `opt-out' consent --- continuing to use a website without active objection to a notice --- is not a clear positive action and as such will not establish a valid legal basis to lay cookies or process data on the basis of consent~\cite{article29workingpartyGuidelinesConsentRegulation2018,cnilcookies,informationcommissionersofficeGuidanceUseCookies2019}. 

As a consequence of the importance of the freely given nature of consent, design matters for legal compliance. Pre-ticked boxes, which require a positive action to opt-out from, are explicitly singled out in the GDPR as an invalid form of consent~\cite[recital 32]{gdpr}. The Court of Justice has recently ruled that they were also not a valid form of consent under the previous law, operational since the mid-90s~\cite{planet49court}. The UK's Information Commissioner's Office further states that ``[a] consent mechanism that emphasises `agree' or `allow' over `reject' or `block' represents a non-compliant approach, as the online service is influencing users towards the `accept' option.'' Similarly, cookie boxes without a `reject' option, or where it is located in a `more information' section or on a third party webpage, are also non-compliant~\cite{informationcommissionersofficeGuidanceUseCookies2019}. One of the CJEU's Advocates General (official impartial advisors to the Court on cases raising new points of law) has emphasised the need that both actions, ``optically in particular, be presented on an equal footing''~\cite[para 66]{planet49ag}. 

Moreover, it must be ``as easy to withdraw as to give consent''~\cite[art 7(3)]{gdpr}. This means if consent was gathered through ``only one mouse-click, swipe of keystroke'', withdrawal must be ``equally as easy'' and ``without detriment'' or ``lowering service levels''~\cite{article29workingpartyGuidelinesConsentRegulation2018}.

An issue of continued contention is the validity of so-called `cookie walls', whereby consent is a prerequisite to accessing a website. While several regulators appear minded to suggest in many or all cases this practice is illegal~\cite{authoriteitpersoonsgegevensHoeLegtAP2019,cnilcookies,europeandataprotectionsupervisorEDPSOpinionProposal,informationcommissionersofficeGuidanceUseCookies2019}, the issue remains unclear~\cite{zuiderveenborgesiusTrackingWallsTakeItOrLeaveIt2017} and the final conclusion will regardless be subject to the ``glacial flow'' of the draft ePrivacy Regulation through the EU's legislative process~\cite{tobinOisin}.

\subsection{Specific and informed consent}
An important aspect of data protection is \emph{purpose limitation}, meaning users must consent in relation to a particular and specific purpose for processing data. They cannot provide \emph{carte blanche} for a data controller to do whatever they like.\footnote{As an unalienable fundamental right, it is impossible for an EU resident to `sign away' their right to effective data protection.} These purposes cannot be inextricably `bundled', so an `accept all' button is only compliant if it is additional to the possibility of specifically consenting to each purpose~\cite{cnilcookies}.

Furthermore, consent is invalid unless all organisations processing this data are specifically named~\cite{cnilcookies,informationcommissionersofficeUpdateReportAdtech2019}. Simply linking to an external list of potential vendors, which may not represent the code being run on the linking webpage, is ``insufficient to provide for free and informed consent''~\cite{informationcommissionersofficeUpdateReportAdtech2019}. Consent should be able to be rejected at the same level as the `accept' button, so having to navigate further to third party websites to reject tracking is non-compliant~\cite{informationcommissionersofficeGuidanceUseCookies2019}. Information required to be provided to data subjects includes certain GDPR--mandated information (including controller contact, processing purposes, legal basis, recipients and sources of data, international transfers, storage period, data rights and rights of complaint, and meaningful information about the logic of significant automated decision-making)~\cite[arts 13--14]{gdpr}, as well as the duration of cookies~\cite{planet49ag,informationcommissionersofficeGuidanceUseCookies2019}.

\subsection{Efficient and timely data protection}
Individuals have the right to `efficient and timely' protection of their data rights, meaning where consent is required, it is required prior to data processing, not subsequently~\cite{article29workingpartyGuidelinesConsentRegulation2018,fashionid}. Cookies must not be set before the user has expressed their affirmative consent. Furthermore, fresh consent is required when new, non-essential cookies are being set by a new third party~\cite{article29workingpartyGuidelinesConsentRegulation2018,informationcommissionersofficeGuidanceUseCookies2019}. The burden is on the data controller to be able to demonstrate that they adhere to data protection law and principles, including that they have valid consent for each individual~\cite[art 5(2)]{gdpr}.

\section{Related Work}
\begin{figure*}[t]
\centering\subcaptionbox{First page\label{qc0}}[0.33\linewidth]{\includegraphics[height = 0.2\textheight]{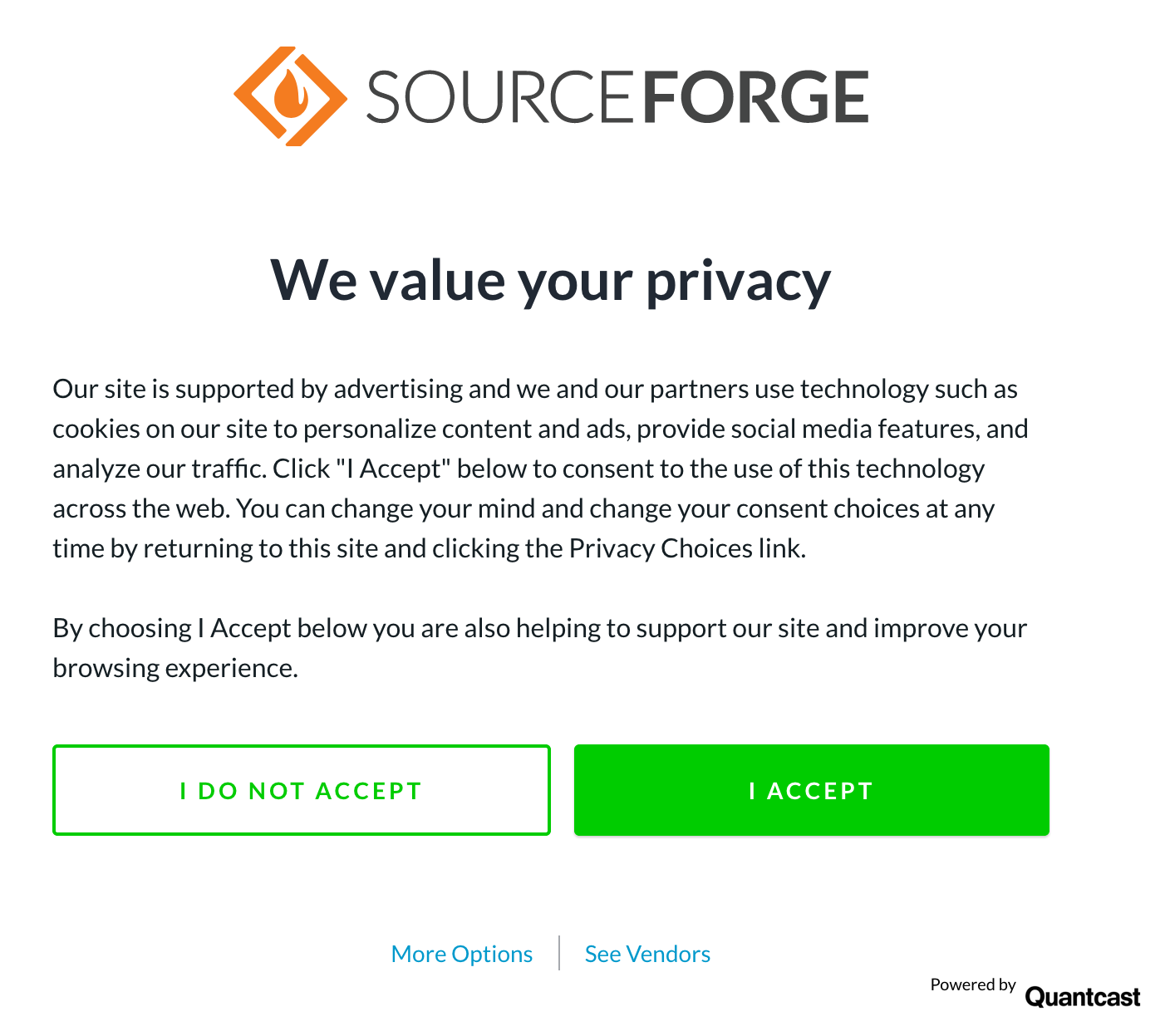}}\subcaptionbox{Categories and purposes\label{qc1}}[0.33\linewidth]{\includegraphics[height = 0.2\textheight]{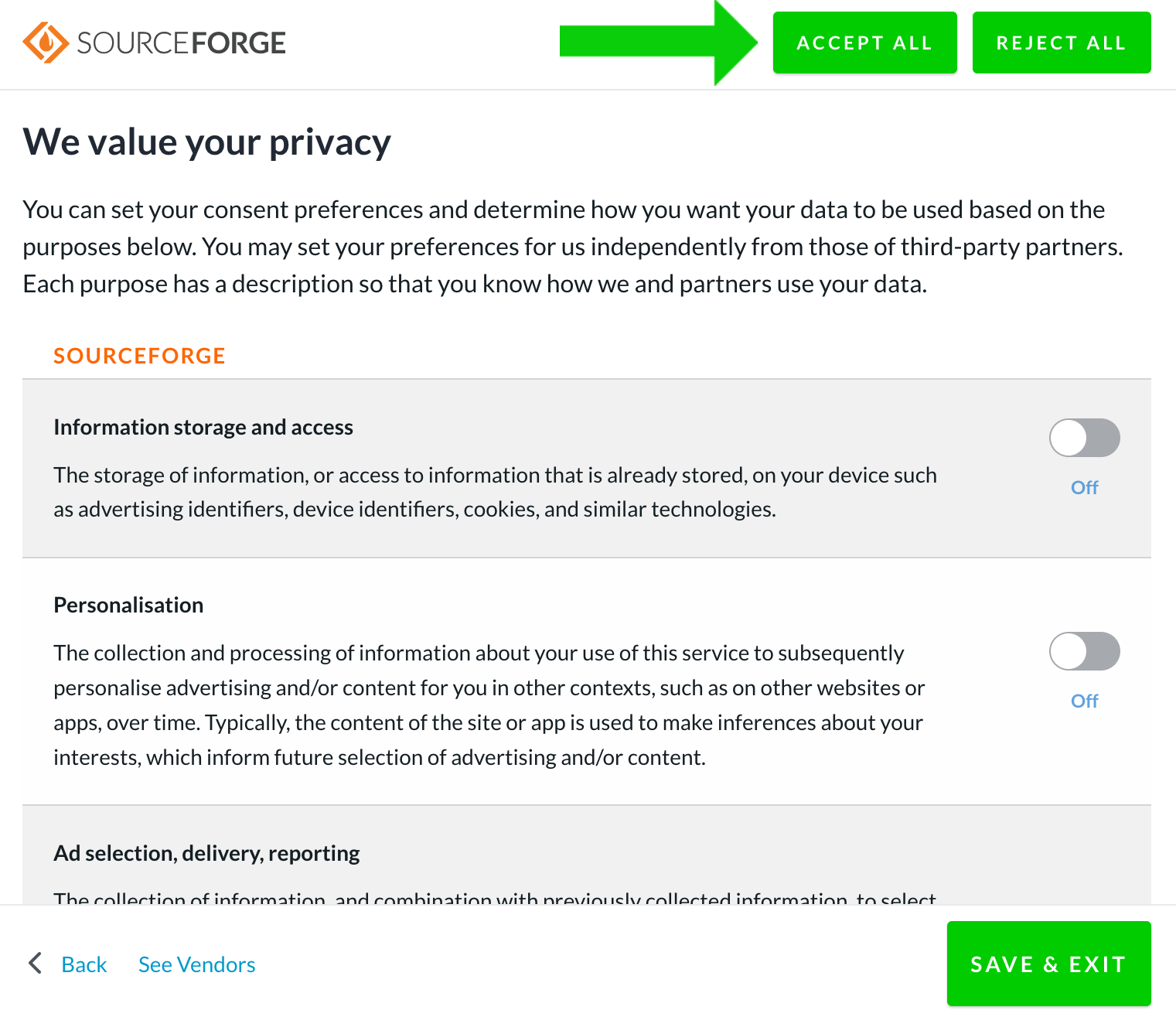}}\subcaptionbox{Vendors/third-parties\label{qc2}}[0.33\linewidth]{\includegraphics[height = 0.2\textheight]{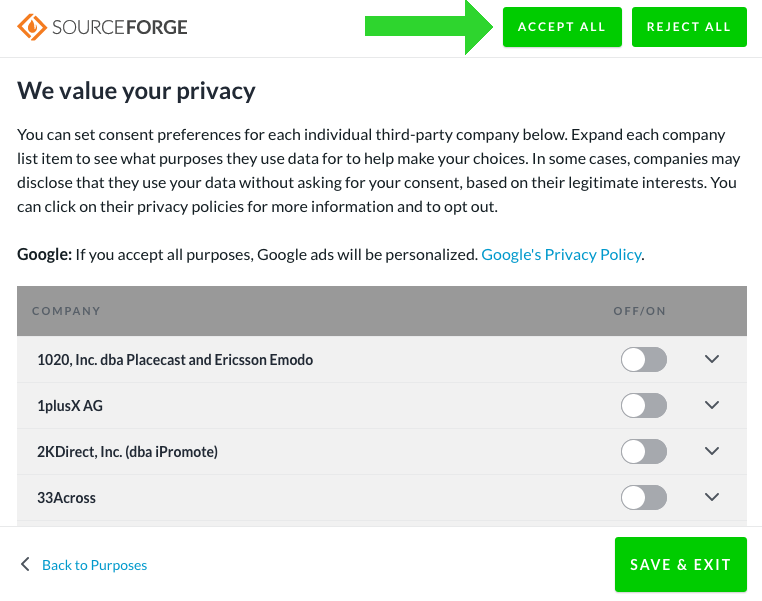}}\caption{The three components of the QuantCast CMP on \url{https://sourceforge.net} as of September 2019.}\label{qc3}
\end{figure*}
\subsection{Notice \& Consent}
The predominant model for communicating information privacy protections to end-users has been notice(/awareness) and consent(/choice).
The interface designs of this model have mostly been privacy policies and opt-in/out interfaces \cite{cranorNecessaryNotSufficient2012}, which legally can be seen as ``pre-formulated declarations of consent'', or ``clickwrap'' contracts~\cite{clifford2019pre}. The usability challenges of these interfaces have seen considerable work across disciplines, largely divisible in studies that establish the shortcomings of interface designs, and studies proposing alternative technologies.
Privacy policy notices are notorious for taking a disproportionate amount of time to go through and require reading comprehension abilities at university level \cite{jensen2004privacy}. 
Privacy policies are rarely read by users \cite{Nissenbaum2011, Obar2018, Vila:2003} prior to using or visiting a site/service. Users have been shown to (almost automatically) consent without viewing them \cite{Acquisti2005, Angulo2011, Meinert2006, McDonald2008, Nissenbaum2011} since they stand in the way of the users' primary goal: accessing the service \cite{Acquisti2005, Angulo2011}. 
This behaviour has been attributed to the users' difficulty understanding how to make meaningful decisions about their privacy preferences; but even in situations where they are made aware of the implications of their decision, they prefer short-term benefits over long-term privacy \cite{Acquisti2005}.
Because of this, control mechanisms of these notices are considered illusory in practice \cite{cate2010limits} --- sometimes having devolved into merely an informational statement rather than an interactive control panel.

The perceived ineffectiveness of this approach has given rise to a number of design alternatives (for an overview of the entire design space, see \cite{schaub2015design}).
Gage Kelley et al. proposed standardised ``nutrition label'' notices with icons representing the type of data collected and how it is used, and showed how it helped users find information more quickly and accurately \cite{kelley2009nutrition}.
Reeder et al. developed an interactive matrix visualisation called Expandable Grid which shows a colour-coded overview of a policy that can be expanded for more detail \cite{reeder2008expandable}.
The Platform for Privacy Preferences (P3P) was an involved attempt to help automate some of this process by building a machine-readable language for expressing website privacy policies which could then interface with user agents, such as the browser or other privacy applications \cite{cranor2002web}.
While it was implemented by Microsoft for Internet Explorer and Edge, P3P never achieved widespread adoption, partly because its comprehensiveness was seen as too complex for regular website owners to apply but also because there was no regulatory or political impetus to force browser vendors to use it.

The majority of studies around notice and consent have focused on how well the interface design helps users make informed decisions.
This paper focuses more on the legal \textit{quality} of the consent that is collected.

\subsection{Dark patterns}
Interface designs that try to guide end-users into desired behaviour through malicious interaction flows are referred to as ``dark patterns'' \cite{gray2018dark}. 
As a phenomenon they are part of the larger research agenda around persuasive design \cite{fogg2009behavior} and nudging \cite{Acquisti2017,thaler2009nudge}.
The practice of dark patterns for privacy notices --- while only sometimes discussed under this moniker in HCI and privacy literature~\cite{bosch2016tales,conti2010malicious,gray2018dark,mathur2019dark} --- is extensively reported on by consumer protection organisations \cite{forbrukerradet}, white papers \cite{edps}, and popular press \cite{singernyt} (for an excellent overview, see the Norwegian Forbrukerrådet document ``Deceived by Design'' \cite{forbrukerradet}).
Its infamy has led the European Union and data protection officers to specifically highlight certain common dark patterns as non-compliant examples of the GDPR in its advisory documents such as privacy intrusive default settings, hiding away privacy-friendly choices and requiring more effort from the user to select it, illusory or take-it-or-leave-it choices, etc.
Senators from the United States have recently introduced a draft bill specifically aimed at outlawing such practices, stating that it should be prohibited for any large online operator to ``design, modify, or manipulate a user interface with the purpose or substantial effect of obscuring, subverting, or impairing user autonomy, decision-making, or choice to obtain consent or user data'' \cite{detour}.


Since the submission of this paper, two studies have been released that look specifically at the consent management platforms that have appeared in response to the GDPR.

Utz et al \cite{Utz2019} analysed a random sample of 1,000 CMPs and manually categorised them along different design dimensions (e.g., positioning, size, consent options). They found (among other things) that a minimum of 57.4\% used dark patterns to nudge users to select privacy-unfriendly options, and that 95.8\% provide either no consent choice or confirmation only.
They conducted a follow-up experiment to test the effects of the CMP position, the granularity and nudging of choices, and the technicality of the language and presence of a privacy policy link. 
They demonstrate that positioning the CMP in the lower (left) part of the screen increases interaction rates; users are more likely to accept tracking given a binary choice than when given more granular options; acceptance rate increased from a mere 0.16\% to 83.55\% when options were preselected; and technical language and privacy policies have a minor effect on consent choice.

The work by Matte, Bielova, and Santos \cite{Matte2019} investigates the actual consent signal sent from the CMP to the respective data processors. They detect that 12.3\% of 1,426 sites send a consent signal before the user makes a choice. Semi-automatically reviewing 560 sites reveals that 54\% of them contain at least one violation regarding the way consent is determined, asked, or complied with.

\subsection{Empirical Studies of EU Privacy Regulation}
Various studies have tried to chart the impact of European privacy regulation on the collection and processing of personal data on the web, both within its territorial scope and globally. 
A longitudinal 4-year study of the impact of the revised ePrivacy directive on cookie placement shows that 1) 49\% of websites placed cookies before receiving consent; 2) 28\% of websites did not provide any consent mechanism; and 3) the percentage of websites violating the directive stayed constant over the course of 4 years, indicating the policy to be ineffective \cite{trevisan20194}.

With respect to the GDPR, both industry and academia have been monitoring its effects since being introduced in May 2018.
Degeling et al. \cite{degeling2018we} monitored the prevalence of privacy policies on websites before and after the introduction of the regulation, showing that in some EU member states the number of policies increased by 15.7\% (to a total of 84.5\%), while 72.6\% of sites updated documents they already had.
They estimate that a total of 62.1\% of websites in Europe display a consent notice, an increase of 16\% since shortly before the regulation became enforceable.
Adzerk, an ad tech company, places this percentage considerably lower, at a mere 20.4\% \cite{adzerk}, although their methodology is more restrictive than Degeling et al.'s.
Interestingly, QuantCast, one of the largest CMP providers, also brought out a report stating that over 90\% of users (n=1bn) have consented to data processing~\cite{quantcast}.

Sanchez-Rola et al. \cite{SanchezRola} performed an evaluation of the tracking undertaken by 2,000 high-traffic websites and evaluated how information notices and actual tracking behaviour changed.
They found that the GDPR affected EU and US sites in the same way, that consent management platforms reduced the amount of tracking, but that personal data collection is still ubiquitous: 90\% still made use of cookies that were able to identify individual users.
S{\o}rensen and Sokol~\cite{Sorensen} present a more nuanced picture of the shifts in third-party tracker presence and behaviour, showing a decrease mostly present in private websites, whereas websites hosted by public institutions mostly stayed the same between February and September 2018.
Along the same lines, there exists a difference between EU and non-EU private-sector sites, but little difference in public sites.
Depending on the purpose category the tracker falls into, further distinctions can be made. 
The largest shift was visible in data collection for advertising, and the least in those used for cybersecurity.
Overall, only 151 third-party trackers are used by 1\% or more of the websites, while the remaining long-tail of 968 have a share of less than one percent.

\begin{table*}[t!]
\centering
\begin{tabularx}{\linewidth}{llp{3cm}p{2.3cm}p{2.3cm}p{2.3cm}p{2.3cm}}
  \toprule
CMP & Sites & Median vendors\newline (low./upp. quartiles)  & Explicit/implicit\newline consent & Banner/barrier & Preticked\newline options & Minimum \newline compliance\\ 
  \midrule
Cookiebot & 12.5\% (85) & 104 (61, 232)  & 45/40 & 78/7 & 64 (75.3\%) & 2 (5.6\%)\\ 
  Crownpeak & 12.2\% (83) & 38.5 (18.8, 132.3) & 46/37 & 52/31 & 67 (80.7\%)  & 0 (0\%)\\ 
  OneTrust & 24.3\% (165) & 58 (26.5, 104.5) & 47/118 & 158/7 & 108 (65.4\%) & 3 (1.8\%) \\ 
  QuantCast & 41\% (279) & 542 (542, 542) & 279/0 & 132/147 & 90 (32.3\%)  & 73 (26.2\%)\\ 
  TrustArc & 10\% (68) & 87 (38, 152) & 42/26 & 26/42 & 53 (77.9\%)  & 2 (2.9\%)\\ \midrule
  \textbf{all} & \textbf{680} & \textbf{315 (58, 542)} & \textbf{459/221} & \textbf{446/234} & \textbf{382 (56.2\%)} & \textbf{80 (11.8\%)} \\
   \bottomrule
\end{tabularx}
\caption{Key statistics on scraped CMPs.} 
\label{cmpstats}
\end{table*}

\section{Study 1: Scraping CMP Interface Designs}
Little is known about many aspects of consent management platforms on the Web, particularly around the consent modalities, quality of this consent and related practices found in the field in the European Union. The major five CMP vendors offer a wide range of customisation options for their clients, and so from an identification of the CMP vendor it does not follow that many assumptions can be made about the interface design. To understand the status quo of consent management plaform interface design after the GDPR, we developed a web scraper to collect information about the five most commonly used third-party CMPs in the top 10,000 most-visited websites in the United Kingdom.

While their sophistication varies, surveyed CMPs all share similarities in back-end function. When a user accesses a site, the CMP detects their IP address and checks their cookies or local storage for any previously set consent preferences, and retrieves this data. If this fails, or if the CMP decides their preferences have expired, the user is shown a consent notice, and their response is recorded. This consent status is then passed on to any integrated tag firing rules, ad servers, and real-time bidding platforms the website has employed.

Visually, the CMP interfaces generally consist of three parts: 1) a first page describing the general purpose of the consent pop-up, with bulk consent options (`accept all' and, for some, `reject all') (Fig. \ref{qc0}); 2) a second page with a more detailed description of the different data processing categories or purposes (e.g. personalisation, marketing), the ability to toggle them individually or collectively, and a button to submit the current consent state (Fig. \ref{qc1}); and, 3) a third page with a breakdown of all the vendors for whom the data is collected or with which it is shared, again with the ability to toggle individually or collectively, and a button to save these settings (Fig. \ref{qc2}). Not all deployed CMPs have all parts of these interfaces enabled.

\subsection{Method}
We built a Web scraper to collect data about the CMP's visual elements, interaction design, and text content (e.g. names of data processing categories or vendors). The scraper utilised the Python library \textit{Scrapy}\footnote{\url{https://github.com/scrapy/scrapy}} and JavaScript rendering service \textit{Splash}\footnote{\url{https://github.com/scrapy-plugins/scrapy-splash}}. The variables the scraper collected included the CMP vendor, the notification style (banner, barrier, other), the type of consent (explicit or implicit) and specific user actions counted as consent (consent/visit/navigation/reloading/scrolling/closing the pop-up/clicking the page); the existence of accept and reject all buttons and the minimum number of clicks to make them available; for both vendors and categories/purposes, the existence of lists of these, their extent and descriptions, whether or which are enabled for user control, and their default state(s). 

We ran the scraper from a Danish IP address\footnote{Relevant legislation is harmonised across the EU and so a Danish IP and UK IP are the same jurisdiction for our purposes.} over 3 days in September 2019 over the top 10,000 UK sites according to webtraffic service \emph{Alexa}. 
We throttled our scraper to two concurrent URL requests and no concurrent requests per domain, with a delay of 2 seconds.
We cycled through three different user agents copied from our browsers to make sure the websites treated us as normal visitors, rather than an automated crawler. 
The CMPs the scraper was designed for are third-party services as identified by \emph{Adzerk} in August,\footnote{A company that does server-side ad serving and writes reports about the state of the industry: \url{www.adzerk.com}} which together account for \textasciitilde58\% of the market share: QuantCast, OneTrust, TrustArc, Cookiebot, and Crownpeak.
We targeted UK sites, rather than sites across all EU countries, because the Adzerk report gives us information about the total population of CMPs in the UK market. This allowed us to check that our scraper's sample was representative both in number of CMPs identified and the overall distribution of the five most popular ones.

To determine the presence of a particular CMP, the scraper looked for an identifying HTML element within 5--15 seconds of arriving on the site (depending on the particular CMP and how it injects the pop-up). Data to construct the variables were extracted by querying for elements and attributes, traversing the DOM if no unique indentifiers existed, or accessing globally scoped objects. This data was pushed to a \textit{MongoDB} database. 
Before deployment, the data returned by the scraper was manually validated with 40 randomly selected sites from the list of 10,000 for each of the five CMPs.
The scraper code and dataset will be available as supplementary material alongside the paper.

\subsection{Understanding compliance}

Based on the above section on EU law, we consider three core, measurable conditions that providers will have to meet to be considered legally compliant for the purpose of this study. This serves as a minimum hurdle: meeting these conditions alone will not guarantee compliance with the law, as there are a multitude of aspects and provisions, many of which can only be appropriately assessed qualitatively. However, these are conditions that are testable with the variables from our scraper, and therefore provide a window on the \emph{maximum} level of compliance in the industry today. These conditions are:

\begin{description}
\item[Consent must be explicit] This condition is true if consent is a clear, positive, affirmative act, such as clicking a button, rather than e.g. continuing to navigate a website.
\item[Accepting all is as easy as rejecting all] Consent must be as easy to give as to withdraw/refuse. This condition is met if accepting all takes the same number of clicks as rejecting all, and automatically not met in the case where consent requires no clicks (i.e. Condition 1 is violated)
\item[No pre-ticked boxes] Consent to any vendor or purpose must be through affirmative acts at all granularity. If no non-necessary purposes or vendors are automatically on, this condition is met. 
\end{description}

Factors which could contribute to non-compliance which we did \emph{not} examine include qualitatively considering the information provided (e.g. specificity of purposes, contact details of vendors, provision of the duration of cookies), nor certain visual features such as colour or size or prominence of buttons beyond clicks.

\begin{figure*}[t!]
\centering
\includegraphics[width = \linewidth]{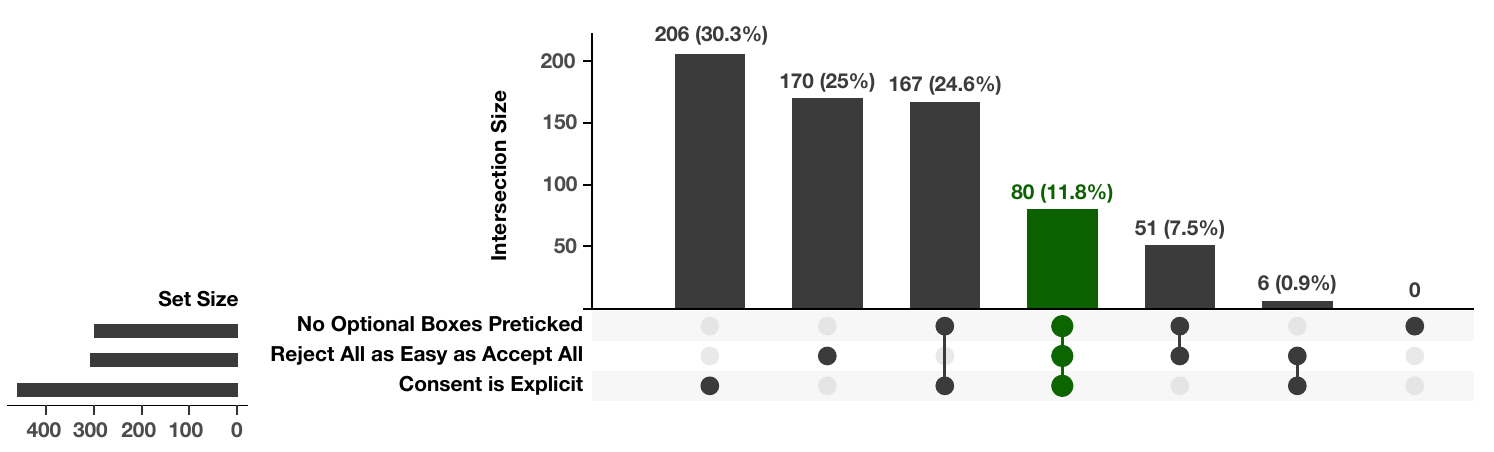}
\caption{UpSet diagram \protect\cite{conwayUpSetRPackageVisualization2017,lexUpSetVisualizationIntersecting2014} of sites by adherence to three core conditions of EU law. Sites meeting all three in \textcolor{diagramgreen}{green}.}
\label{upset}
\end{figure*}

\subsection{Results}

680 (6.8\%) of the top 10,000 UK websites contained a CMP which could be successfully scraped by our tool. According to a survey of the top 10K UK websites in August 2019~\cite{adzerk}, only 20.35\% of the top 10K UK websites are reported to use a CMP (from any vendor). 1191 of those (i.e., 58.52\%) use the top 5 CMPs, which means the 680 instances our scraper captured represents 57.09\% of the total population\footnote{It should be noted that Adzerk's methodology counts CMPs by URL endpoints of the Javascript files and we found during development that websites frequently include inactive CMPs' .js files. This means that Adzerk's statistics are likely inflated with double-counting, and that our survey is consequently more representative than the 57.09\% would indicate.}.

We found that implicit consent is common among these sites (32.5\%). An array of actions that websites count as consent (but which EU law does not) was extracted from their code, such as just visiting the site (16.8\%), navigating within the site (6.2\%), revisiting/refreshing the page (7.6\%), scrolling or clicking on the page (5.3\%) or closing the pop-up or banner (1.6\%). 9\% of sites accepted more than one form of implicit consent. With only a handful of idiosyncratic exemptions all implied consent was found in the use of `banner' rather than `barriers' (a barrier style is in Fig. \ref{qc3}). Within those CMPs exhibiting explicit consent, there was a roughly even split between the use of barriers and banners (50.3\%/49.7\%). Popular CMP implementation wizards still allow their clients to choose implied consent, even when they have already indicated the CMP should check whether the visitor's IP is within the geographical scope of the EU, which should be mutually exclusive. This raises significant questions over adherence with the concept of \emph{data protection by design} in the GDPR.

The vast majority of CMPs make rejecting all tracking substantially more difficult than accepting it. 50.1\% of sites did not have a `reject all' button. Only 12.6\% of sites had a `reject all' button accessible with the same or fewer number of clicks as an `accept all' button. In practice, this means both were accessible on the first page --- an `accept all' button was never buried in a second layer. 74.3\% of reject all buttons were one layer deep, requiring two clicks to press; 0.9\% of them were two layers away, requiring at minimum three.

Furthermore, when users went to amend specific consent settings rather than accept everything, they are often faced with pre-ticked boxes of the type specifically forbidden by the GDPR~\cite[recital 32]{gdpr}. 56.2\% of sites pre-ticked optional vendors or purposes/categories, with 54.1\% of sites pre-ticking optional purposes, 32.3\% pre-ticking optional categories, and 30.3\% pre-ticking both. Our scraper was detecting visual status rather than functional status---we do not know the impact on toggling on or off vendors or categories beyond what the CMP tells the user is happening (Matte et al.'s \cite{Matte2019} findings indicate 7.7\% of CMPs ignore the consent signal submitted by the user).

Sites relied on a large number of third party trackers, which would take a prohibitively long time for users to inform themselves about clearly. Out of the 85.4\% of sites that did list vendors (e.g. third party trackers) within the CMP, there was a median number of 315 vendors (low. quartile 58, upp. quartile 542). Different CMP vendors have different average numbers of vendors, with the highest being QuantCast at 542 (see Table \ref{cmpstats}). 75\% of sites had over 58 vendors. 76.47\% of sites provide some descriptions of their vendors. The mean total length of these descriptions per site is 7985 words: roughly 31.9 minutes of reading for the average 250 words-per-minute reader, not counting interaction time to e.g. unfold collapsed boxes or navigating to and reading specific privacy policies of a  vendor.

As discussed, we consider that a site is minimally compliant if it has no optional boxes pre-ticked, if rejection is as easy as acceptance, and if consent is explicit. Only 11.8\% of sites met these basic requirements. The interaction between the requirements is shown in Figure~\ref{upset}. This varied significantly by CMP vendor --- as shown in Table~\ref{cmpstats}, only Quantcast has a non-negligible number of CMPs that we consider minimally compliant (26.2\%), with Crownpeak having zero (that we found). This can largely be explained by the non-existence of implicit consent in QuantCast CMPs and their lower levels of pre-ticked boxes.

\subsection{Interim Discussion}

Given that all vendors (with the exception of Crownpeak) have examples in the wild of minimally compliant CMPs, it is unclear whether non-compliance is a practical result of sites configuring it in a non-compliant manner, being encouraged to do so by the CMP vendors or, in some cases, running older CMPs without updating them in light of the more publicised nature of the law.\footnote{Note that the recent judgement from the European Court of Justice clarified that these requirements have been part of EU law since 2012, rather than just since the GDPR~\cite{planet49court}} Whatever the practical reasons, 11.8\% is an extraordinarily low number for seemingly market-leading CMP vendors, and suggests an urgent role for data protection authorities to take action to ensure only correct configurations are permitted.

The dataset in this study will be available to other researchers, and we welcome further research into, for example, the scraped text content of the CMPs, as the 11.8\% in this study is a maximum value that is likely to only decrease on consideration of further aspects of the law which are harder to assess in a formulaic manner.

\subsection{Limitations}
Although we manually validated the scraper, we cannot guarantee that there are no false negatives or false positives in our dataset.
Because these CMPs are dynamically rendered via JavaScript, determining whether the state of the DOM scraped is the final one is tricky (further complicated by the fact that Scrapy's engine runs on ECMAScript 2015 making tools to deal with asynchronous execution, such as \emph{async}/\emph{await}, unavailable). We hardcoded a waiting time of 5-15 seconds between loading the site and scraping the content which should be more than sufficient, but there might be exceptions.
The CMP might be customised either by the company or the website owner, thwarting the automated way we identify the presence of elements.
Legacy implementations, either from various iterations over the years or because the company has been sold multiple times, also introduced branches in the CMP code we might have missed.
While we did our best to identify and work around elements of the CMPs designed to obfuscate their function and prevent automation, deliberate changes to data retrieval are often used to foil research for those studying APIs~\cite{brunsAPIcalypseSocialMedia2019,bucherObjectsIntenseFeeling2013}, and such practices seem likely in this domain also to protect against potential automated regulatory scrutiny.
\section{Study 2: effects of designs on answers}
The goal of the second study was to establish if, and to what extent, certain CMP designs affect the consent answer given by users. 
We were interested in non-compliant designs that are very prevalent, or designs that are not yet described as non-compliant by the applicable regulation.
We conducted two field experiments to establish the effects on user behaviour and consent rate of 1) barrier and banner notifications; 2) equal and unequal prominence of accept all and reject all options on the first page; and 3) the level granularity of consent options on the first page (bulk, purposes, vendors).



\subsection{Method}

\subsubsection{Design}
\begin{figure*}[t]
\centering
\begin{subfigure}{1\linewidth}
    \centering
    \subcaptionbox{
\label{banar}}[0.45\linewidth]{\includegraphics[width=0.46\textwidth]{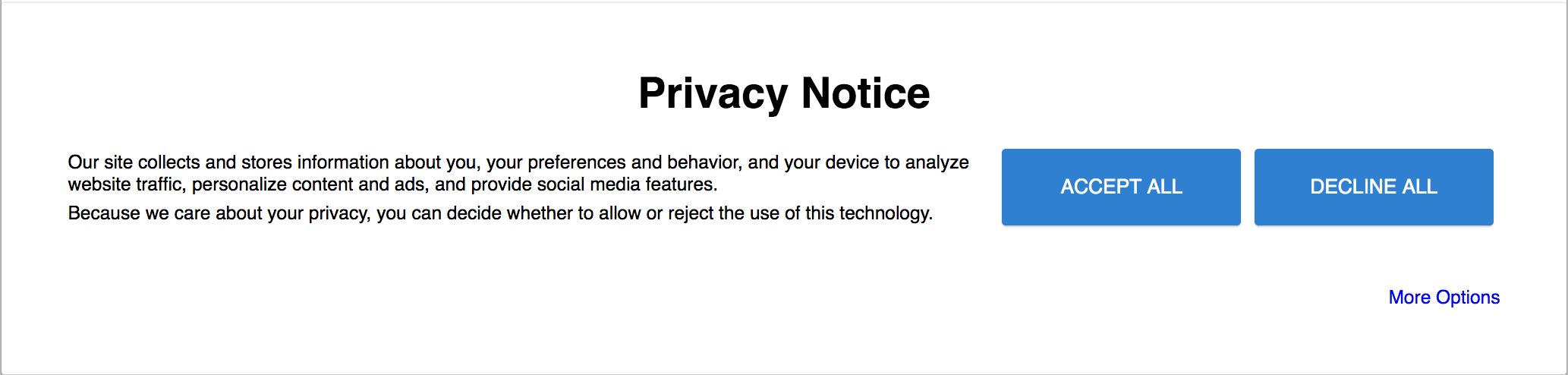}}
    \subcaptionbox{
\label{barar}}[0.17\linewidth]{\includegraphics[width=0.17\textwidth]{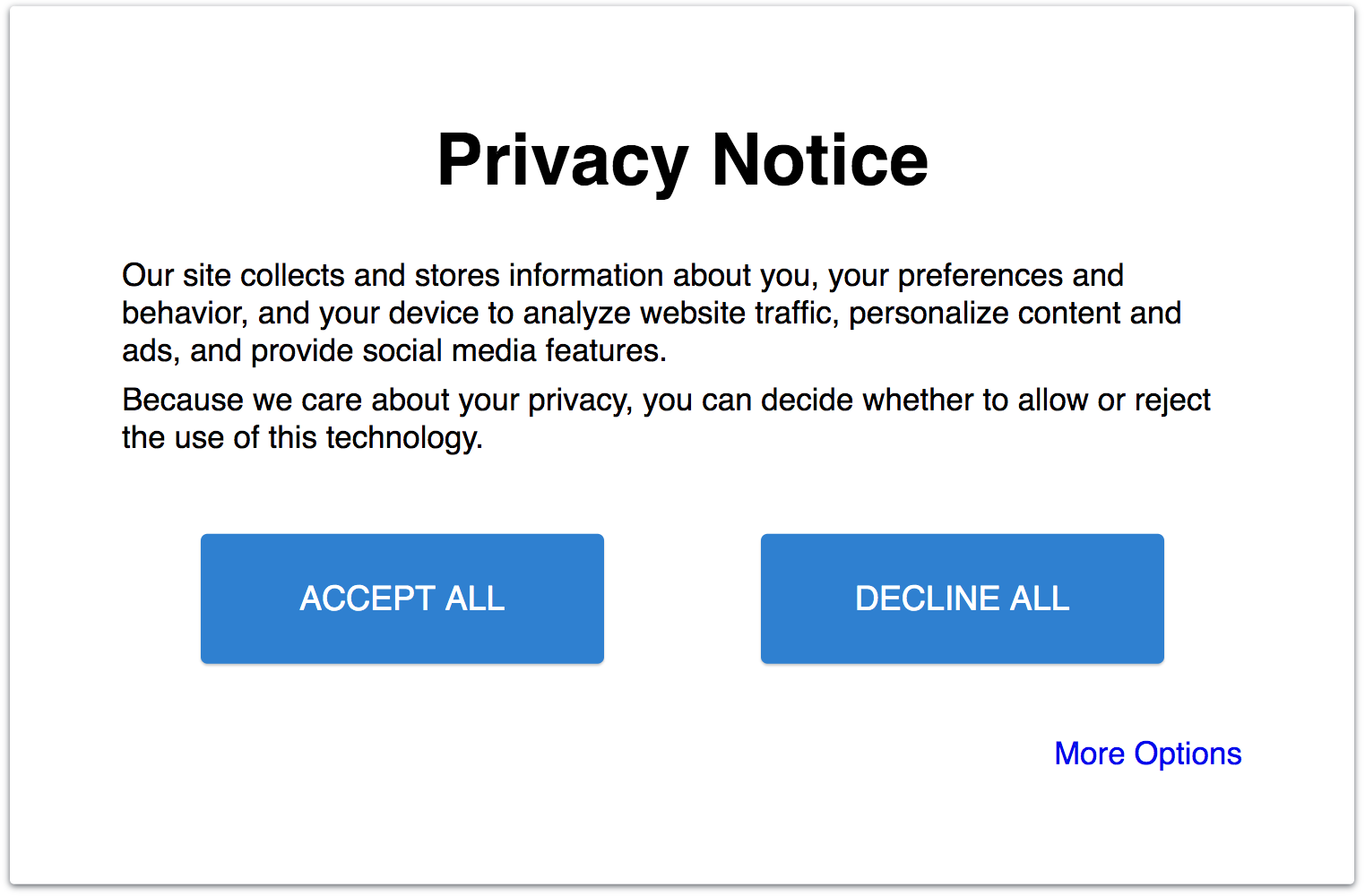}}
    \subcaptionbox{
\label{bulk}}[0.19\linewidth]{\includegraphics[width=0.19\textwidth]{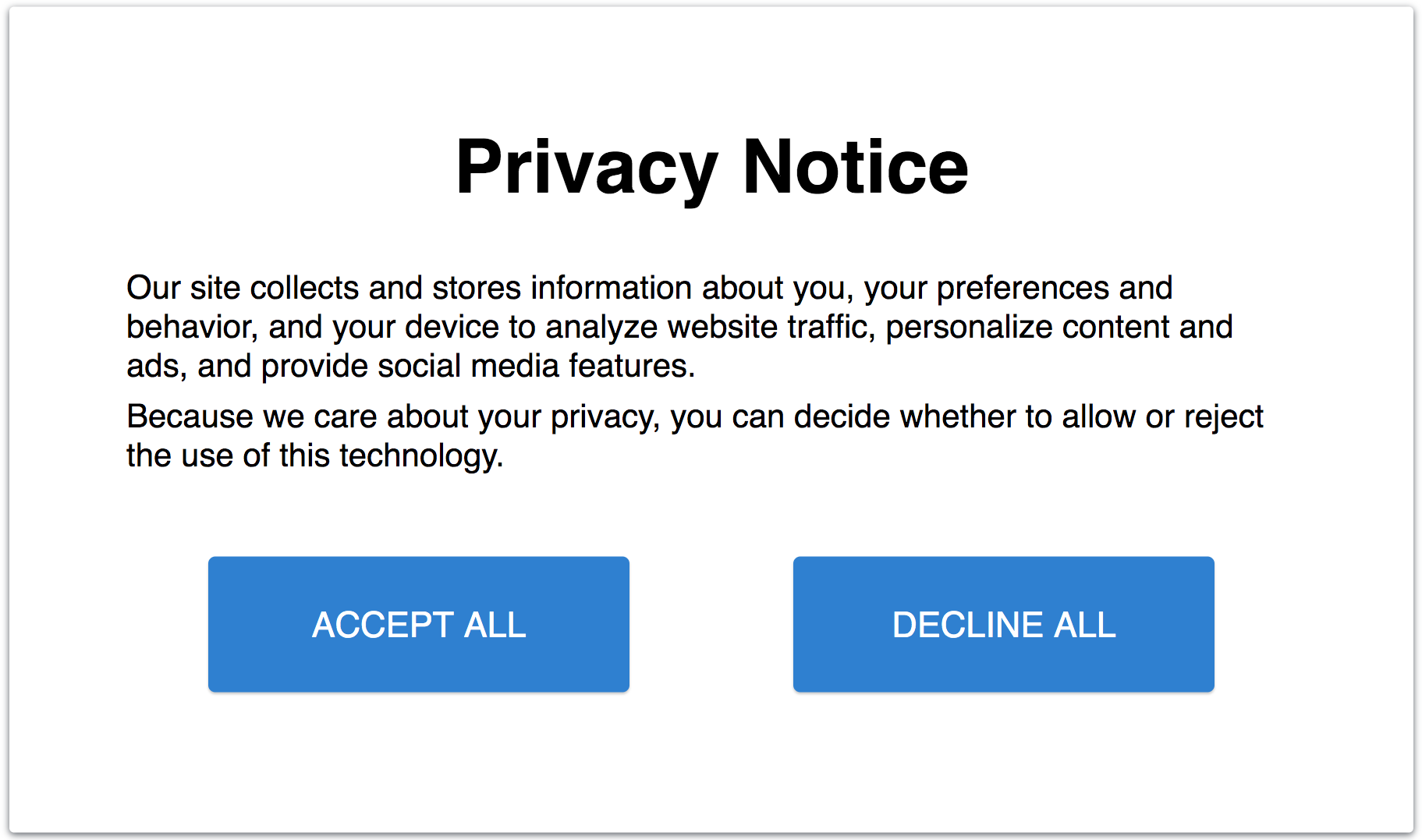}}
    \subcaptionbox{
\label{bp}}[0.16\linewidth]{\includegraphics[width=0.156\textwidth]{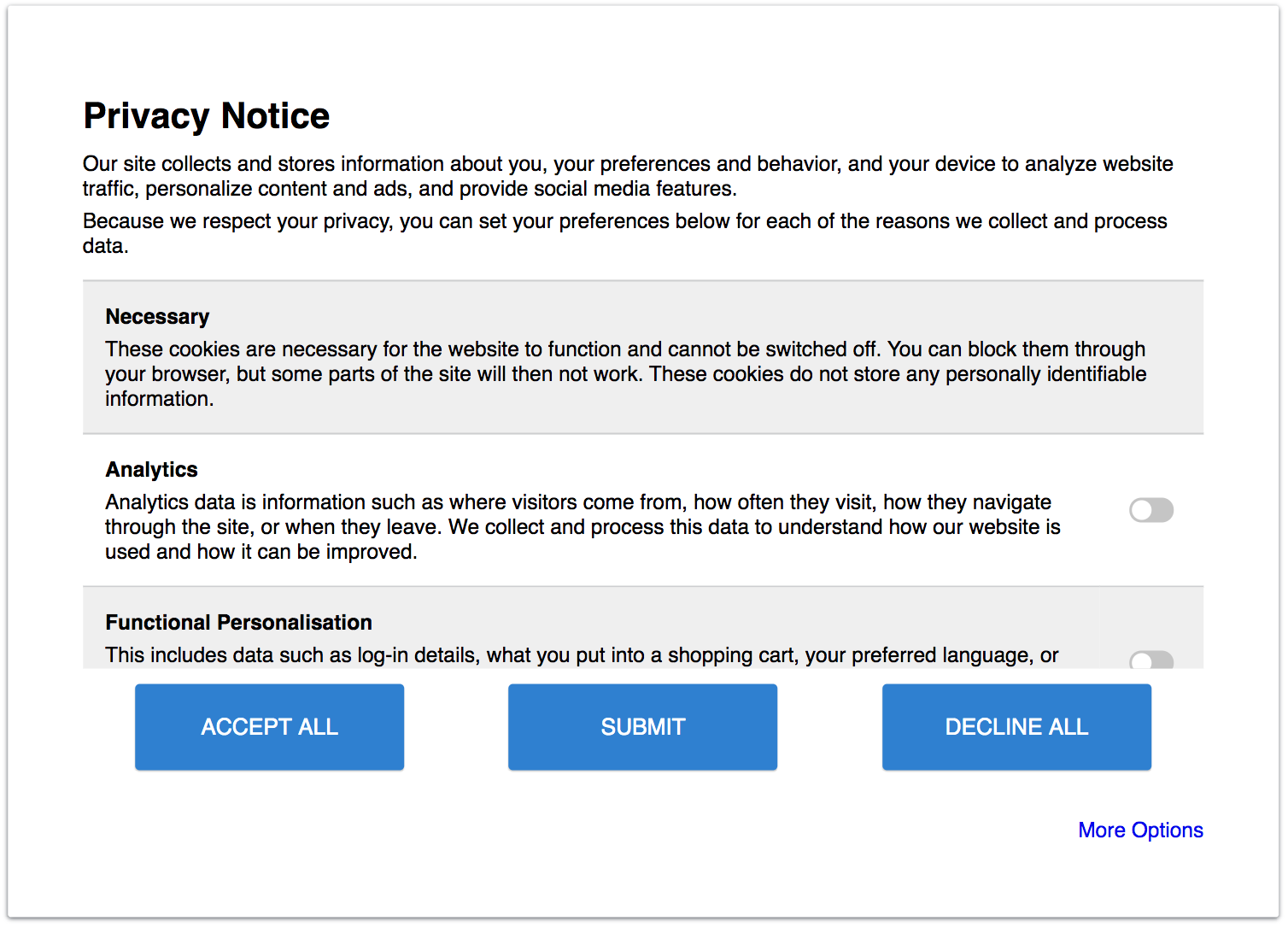}}
\end{subfigure}
\begin{subfigure}{1\linewidth}
    \centering
    \subcaptionbox{
\label{banar2}}[0.45\linewidth]{\includegraphics[width=0.46\textwidth]{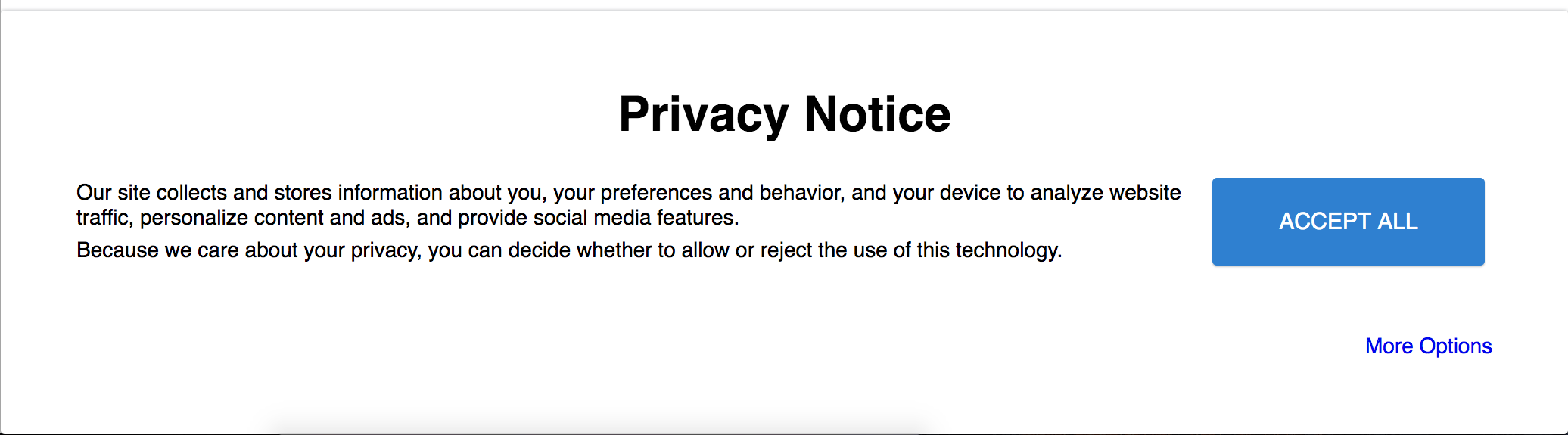}}
    \subcaptionbox{
\label{banar3}}[0.19\linewidth]{\includegraphics[width=0.19\textwidth]{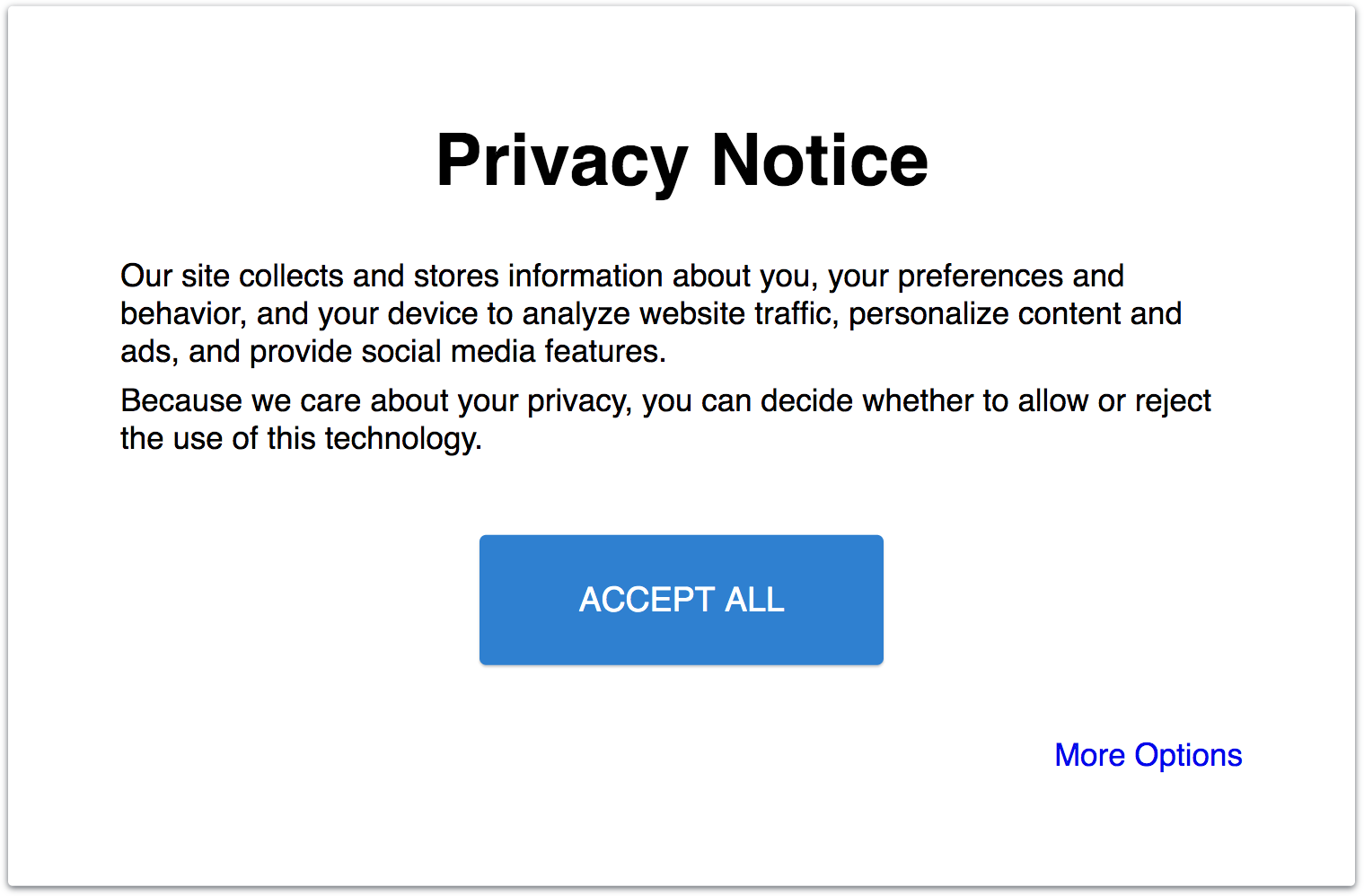}}
    \subcaptionbox{
\label{banar4}}[0.17\linewidth]{\includegraphics[width=0.173\textwidth]{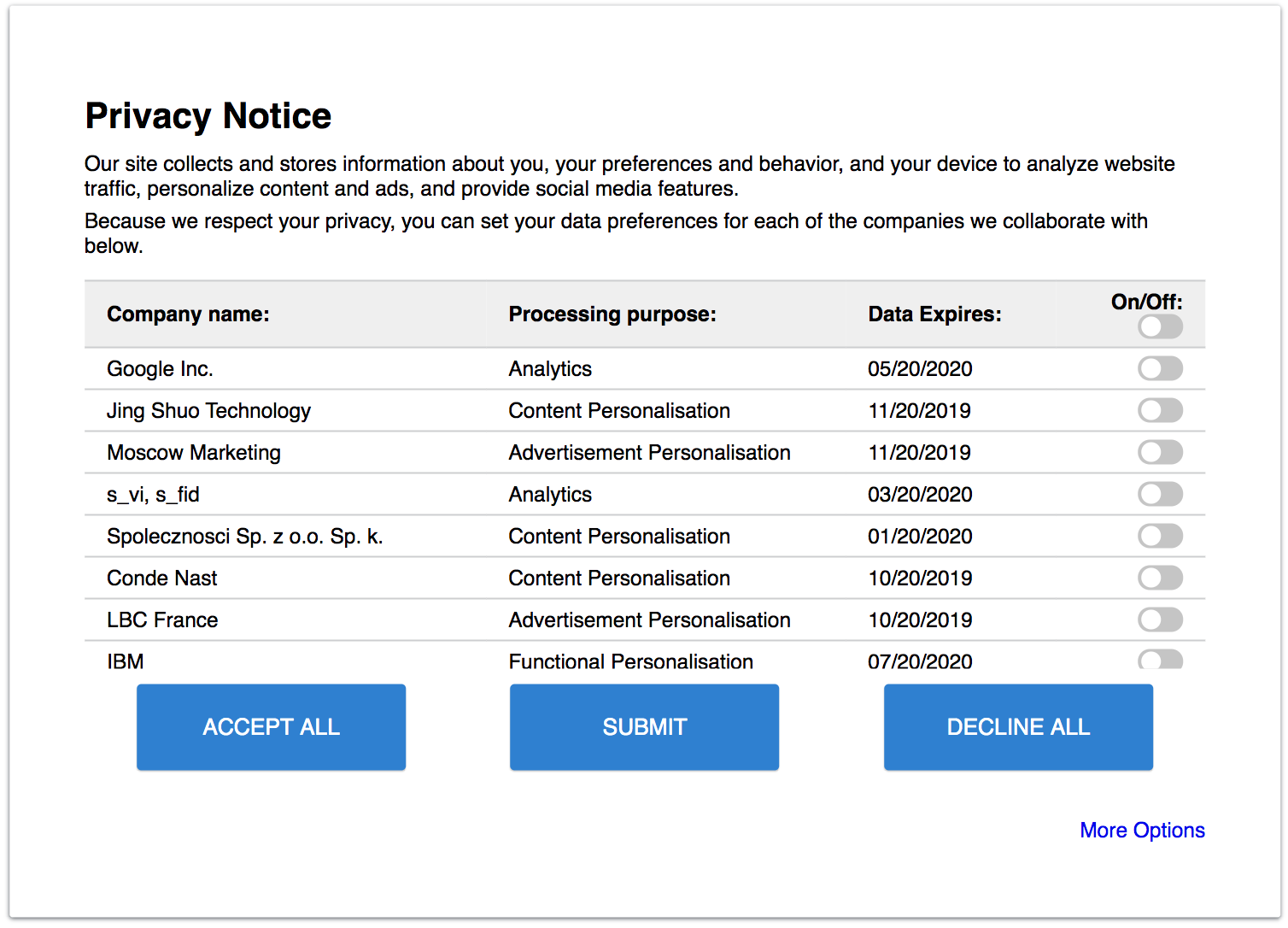}}
    \subcaptionbox{
\label{banar5}}[0.17\linewidth]{\includegraphics[width=0.185\textwidth]{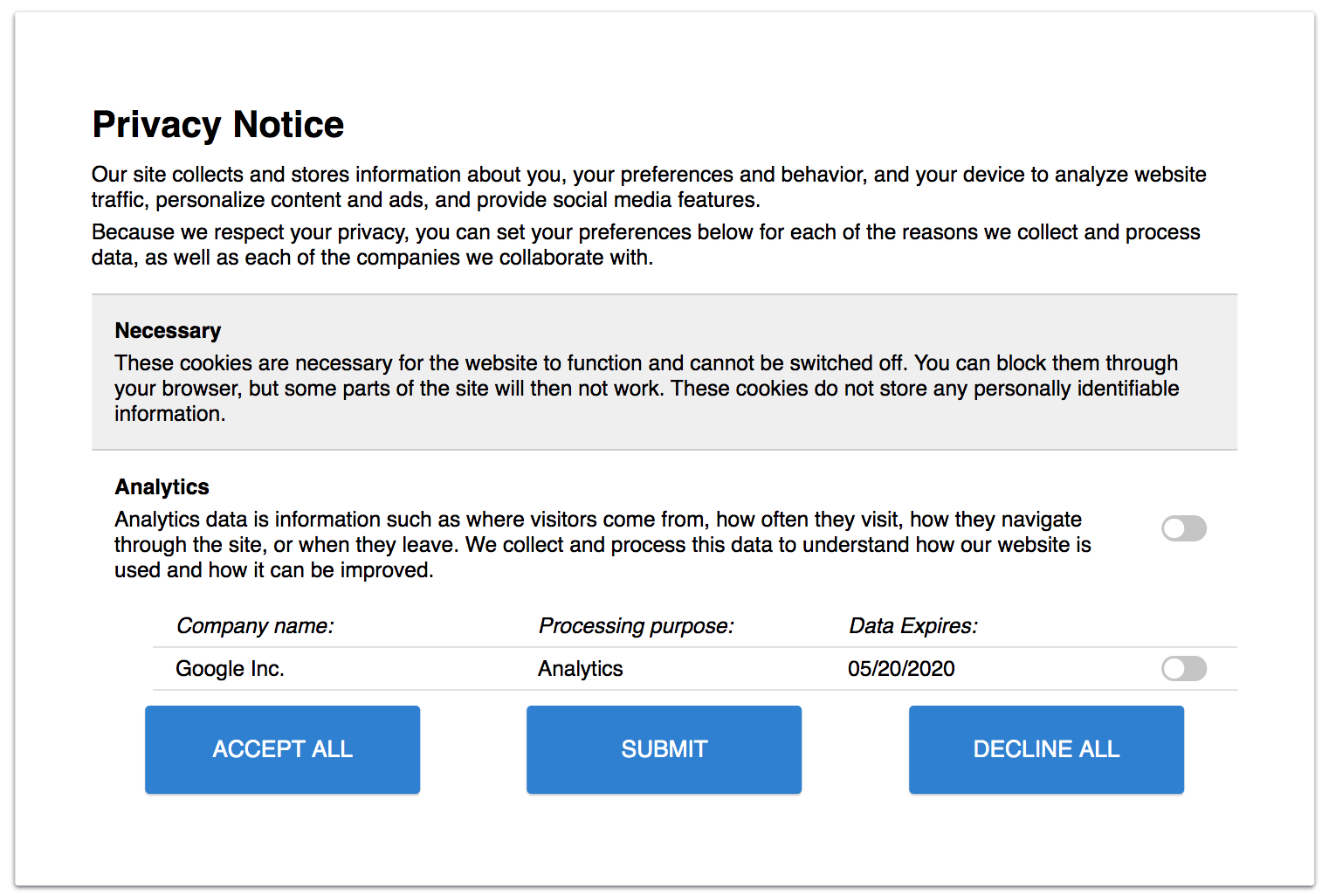}}
\end{subfigure}
\caption{The 8 interface conditions: (a) Banner / Accept + Reject; (b) Barrier / Accept + Reject; (c) Bulk; (d) Bulk + Purposes; (e) Banner / Accept; (f) Barrier / Accept; (g) Bulk + Vendors; (h) Bulk + Purposes + Vendors.}
\label{interfaces}
\end{figure*}

The study consisted of two counter-balanced experiments, evaluating a total of 8 different interfaces (see Figure \ref{interfaces}).

\textit{Experiment 1} used a [2x2] latin square, within-subjects, repeated measures design.
The independent variables were the \textsc{notification style} (\textit{Barrier; Banner}) and \textsc{bulk consent buttons} (\textit{Accept all+Reject all; Accept all}).
The primary dependent variable was the \textsc{consent answer} (\textit{Accept all; Reject all; Submit default; Submit personalised}).

\textit{Experiment 2} used a [1x4] latin square, within-subjects, repeated measures design.
The independent variable was the \textsc{consent granularity} (\textit{Bulk; Bulk+Purposes; Bulk+Vendors; Bulk+Purposes+Vendors}) on the first page of the notification.
The primary dependent variable was the \textsc{consent answer} (\textit{Accept all; Reject all; Submit default; Submit personalised}).

\subsubsection{Participants}
A total of 40 participants successfully finished both experiments, with a  mean age of 26.1 and standard deviation of 8.6\footnote{Age was reported using brackets of ten years so we are unable to report the exact range; the answers were assumed to be normally distributed to calculate the mean.}
The majority, 30, had a university degree (17 Bachelor, 12 Master, 1 Doctorate).
Seven had some college credit but no degree, and three a highschool diploma.
28 participants were currently studying and 12 were employed full-time.
All participants were residing in the United States for the duration of the study, and did not travel to the EU.
We selected this sample to prevent the confounding effects of real CMPs, which would have popped-up on top of our injected one if the participants were in the EU and thus in the regulatory scope of the GDPR.
Four participants lived in the EU in the past five years, meaning they might already be familiar with pop-ups from the ePrivacy directive. 
All participants used Google Chrome as their main browser.

Participants were recruited through one of the author's personal network and a university mailing list.
They were offered \$50 upon completion of the study, and an additional \$10 if they successfully recommended others.

\subsubsection{Apparatus and Materials}
The materials and apparatus of this study include a pre-study survey, a browser extension, and a post-study survey.

The pre-study survey consisted of 11 questions designed to gather demographic information (age, employment status, highest degree obtained, country of residence), check whether the participants met the study criteria (devices used to browse the web, main browser, travelling to the EU during the study), and acquire their informed consent.

To expose the participants to the different interface designs in a controlled yet ecologically realistic context, we developed a browser extension that injects different pop-ups into any website that participants would visit during their normal daily browsing (available as open-source after publication).
The designs of the eight interfaces (i.e., conditions) were inspired by the designs of the top five Consent Management Platforms also used for the scraper study: QuantCast, OneTrust, TrustArc, Cookiebot, and Crownpeak.

All the text, data processing purposes, and vendor names were created by synthesising those commonly used by a random selection of those CMPS in the top 500 Alexa websites in the UK.
The data processing purposes are a combination of the options that the five CMPs give to website owners when they create their own pop-up, or the purposes those websites came up with themselves.
The vendor names were copied from existing websites, and picked to represent one of four categories: well-known companies (e.g, ``Yahoo!''), foreign companies with English names (e.g., ``Beijing Interactive Marketing''), foreign companies with non-English names (e.g., ``Programatica de publicidad S.L.''), and gibberish names (e.g., ``s\_vi\_bikx7Becalgbkjkxx'').

The extension used the open-source JavaScript database PouchDB to store the participants' interactions with the interfaces locally, which was synchronised with a CouchDB instance running on OpenStack over an SSL encrypted connection. 

The post-study survey consisted of four questions asking the participants to reflect on their general pop-up answering strategy, showed them a visualisation of their actual answers, and asked them to describe how well those answers fit their ideal preferences.

\subsubsection{Procedure}
A recruitment email was sent to potential participants asking them to join a study about web-tracker activity in the United States compared to the European Union, and answer the pre-study survey.
Once approved, the participants were assigned and emailed a participant number and a link to the Chrome extension on the Chrome Web Store.
After installing the extension, a welcome screen automatically appeared asking the participants to fill in their assigned number.
This connected the installation to the participant number in the CouchDB database, where each participant was matched to a pre-determined experiment and condition order.
Once the extension was successfully activated, a pop-up appeared notifying the participants the experiment had started.
To train the participants and homogenise their understanding of the CMPs they received an additional email that informed them they might sometimes see consent pop-ups (ostensibly when they were shown the European version of a website instead of the US equivalent), explained how those pop-ups worked, and instructed them to answer the pop-ups according to their preferences. 

The extension injected a pop-up every fourth url visited -- including navigations on the same page, excluding automatic redirects or urls for which an answer was already recorded -- to approximate the realistic frequency with which consent pop-ups are currently shown\footnote{Based on Adzerk's Ad-Tech Insights report: \cite{adzerk}}.
Each interface condition was repeated four times, requiring the participants to answer sixteen pop-ups per experiment.
All interactions with the pop-up were recorded and timestamped: clicking on the elements, toggling purposes or vendors, scrolling the lists, navigating back and forth between the pages, submitting a consent response.
Interfaces which were not interacted with were re-appended to the list of conditions and shown again for a maximum of five times, after which it was recorded as ``not answered'' (similar to a participant clicking or scrolling through the interface without providing a consent response).
Once all conditions of the first experiment were answered, the participant progressed to the second experiment.

After completing both experiments, the participants were notified by email that the study was finished, informed that they could uninstall the extension, and asked to complete the post-study survey.
The completion time of the experiment ranged from four days to three weeks, depending on how many unique urls the participant visited per day (e.g., some participants mostly visited the same websites, some went on holiday during the experiment, some installed the extension on their secondary device and only used it a couple days per week).

\subsubsection{Data analysis}
Although originally 48 participants finished the experiment, we removed eight of them because they mentioned in the survey that their answers were affected by the study (e.g., some participants said they chose ``accept all'' because they wanted to give more data, despite the instructions they received).
To analyse the effects of interface design on consent answers we created a linear regression model with fixed effects; we treated the participants as a factor to account for their (assumed) stable privacy preferences.

\subsection{Results}
\subsubsection{General interaction patterns}
Of the four possible consent choices -- accept all, reject all, submit preferences, no answer -- the vast majority of answers submitted by participants was through the bulk options (89.3\%), with a skew towards accepting: 55.2\% (707) accept all versus 34.1\% (437) reject all.
Just 9.7\% (124) of answers represent the ``submit preference'' option, and 0.9\% (12) were ``no answer'' (but recorded interactions).
Of those 124 ``submit'' answers, merely 21 -- given by 6 participants -- were personalised answers, instead of submitting the default status (all toggled off).
Four of those 21 were personalised by clicking the ``Toggle All'' button, which means \textit{only 17 answers out of 1280 (1.3\%) represent a participant consenting to a specific selection of purposes or vendors.}
Whether this is because users are unable to make such decisions, are not interested in that level of detail, or fatigued by the form and frequency of the question is unclear; but it does indicate that users' consent is rarely empirically as "specific" as the GDPR requires it to be.
It does not follow that specific controls should therefore be removed, but rather that such specificity could be distributed to other actors invited by the user (e.g., browser agents, consent predictors, a knowledgeable friend).

Almost all interactions (93.1\%)
were limited to the first page of the pop-up the participants were exposed to. 
Seven out of eight interfaces had a ``more options'' link to navigate to a second or third page for more information and granular consent choices, but this was clicked only 88 times (6.9\%).
When participants were exposed to a scrollable list of data collection purposes or vendors on the first or subsequent pages (560 occasions), they ignored it 68.6\% (384) of the time.
Of the 176 instances they did scroll, 21.6\% (38) were between 0 and 25 percent of the list, and 64.2\% (113) between 75 and 100 percent.
In other words, anything not immediately visible to the user, anything requiring interaction to access, might as well not exist.

\subsubsection{Notification style}
The validity of one design element still under discussion by policy makers is that of the notification style \cite{zuiderveenborgesiusTrackingWallsTakeItOrLeaveIt2017}: a barrier in the middle of the screen which prevents the user from interacting with the website until a response is recorded, or a banner stretching the width of the screen that does not block access to the information . 

We found that \textit{notification style did not affect the consent rate of participants}.
Two simple linear regressions were calculated to investigate the relationship between the answer given (accept or not) and notification style (banner or barrier). 
The first, comparing \textsc{Barrier} to \textsc{Banner} with both the Accept and Reject button, did not find a regression line at all (F(1,279) = 0.000, p = 1).
The second, comparing \textsc{Barrier} to \textsc{Banner} with just the Accept button, found a non-significant relationship (p = 0.702), with a slope coefficient of 0.013 (95\% CI min and max of $-$0.052 and 0.077 respectively) and an R\textsuperscript{2} of 0.001.

While there was no difference in acceptance rate when participants actually answered the pop-up, the banner notification was ignored 3.6 times more often than the barrier.
For this statistic we considered any pop-up that was not interacted with, but which had a time difference of at least 3 seconds between being injected and the tab being closed, as ``ignored''; 133 of such instances were found, with only 21.1\% (28) for the barrier and 78.9\% (106) for the banner.



\subsubsection{Button prominence}
Data from our scraper indicates `accept all' and `reject all' buttons are not displayed with equal prominence: only a mere 12.6\% of sites show both on the same page.
Such unequal prominence of consent options is already considered non-compliant with the GDPR~\cite{planet49ag, informationcommissionersofficeGuidanceUseCookies2019} because it is expected they affect consent answers, but the severity of its impact is unknown.

We found that \textit{removing the `reject all' button from the first page increased the probability of consent by 22$-$23 percentage points}.
We calculated two simple linear regressions to analyse the relationship between the answer given (accept or not) and the consent options on the first page (accept and reject, or just accept).
The first, comparing \textsc{Accept all + Reject all} to \textsc{Accept all} for the barrier notification, found a strong positive linear relationship between the two. The significant (p < 0.001) slope coefficient for the consent answer was 0.220, meaning the accept rate increased on average by 22.0 percentage points when the reject all button was removed from the first page. 
The 95\% CI had a minimum and maximum of 0.149 and 0.290 respectively.
The R\textsuperscript{2} was 0.117, so 11.7\% of the variation in answers for the barrier notification can be explained by the changing prominence of the buttons.

The second regression compared \textsc{Accept all + Reject all} to \textsc{Accept all} for banner notifications and found a similarly strong, positive linear relationship between the button prominence and answer given.
The significant (p < 0.001) slope coefficient was 0.231, meaning the accept rate increased on average by 23.1 percentage points when the reject all button was removed from the first page.
The 95\% CI had a minimum and maximum of 0.163 and 0.230 respectively.
The R\textsuperscript{2} was 0.135, so 13.5\% of the variation in answers for the banner notification can be explained by the changing prominence of the buttons.

\subsubsection{Level of granularity}
The most common order in which consent options are displayed is bulk first, followed by the data collection purposes on the second page and the vendors on the third page, or some combination of those two on the same page.

We found that \textit{displaying more granular consent choices on the first page decreased the probability of consent by 8$-$20 percentage points}. 
We calculated a simple linear regression to compare a \textsc{Bulk} only interface to an interface that combined \textsc{Bulk + Purposes}; \textsc{Bulk + Vendors}; and \textsc{Bulk + Purposes + Vendors} on the same page. 
We found a significant (p < 0.01) negative relationship between all increases in the level of granularity of consent options and the answer given, with different strengths depending on the kind of options that were available.
As illustrated by Table \ref{granularity}, showing just the vendors affected the acceptance rate the most ($-$0.200), whereas just the purposes ($-$0.088) and the combination of vendors and purposes ($-$0.119) were closer together but still lower than the baseline interface with just bulk options.
Along the same lines, the 95\% CIs overlap most between \textsc{Purposes} and \textsc{Purposes + Vendors} and only a little with \textsc{Vendors}. 


\begin{table}[!htbp] \centering 
  \caption{Level of granularity on the first page, with bulk consent as the reference} 
  \label{granularity} 
\begin{tabular}{@{\extracolsep{1pt}}lcc} 
\\[-1.8ex]\hline 
\hline \\[-1.8ex] 
 & \multicolumn{1}{c}{\textit{Dependent variable:}} &  {\textit{95\% CI:}} \\
\cline{2-3}
\\[-1.8ex] & `accept all' clicked  & lower : upper \\ 
\hline \\[-1.8ex] 
 Bulk + Purposes & $-$0.088$^{**}$ & $-$0.151 : $-$0.024 \\ 
  & (0.032) \\ 
 Bulk + Vendors & $-$0.200$^{***}$ & $-$0.263 : $-$0.137\\ 
  & (0.032) \\ 
 Bulk + Purposes & $-$0.119$^{***}$ & $-$0.182 : $-$0.056 \\ 
  + Vendors & (0.032) \\ 
\hline \\[-1.8ex] 
Observations & 640 \\ 
R$^{2}$ & 0.062 \\ 
F Statistic & 13.210$^{***}$ (df = 3; 597) \\ 
\hline 
\hline \\[-1.8ex] 
\textit{Note:}  & \multicolumn{2}{r}{$^{*}$p$<$0.1; $^{**}$p$<$0.01; $^{***}$p$<$0.001} \\ 
\end{tabular} 
\end{table}

\subsubsection{Participant Strategies and Behaviour Patterns}
While the experimental data suggests how different designs affect how ``freely given'' the consent answers of participants are, it does not provide information about how those answer relate to their preferred privacy settings.
In a post-study survey, we requested participants to describe their overall answering strategy, showed them a visualisation of their actual behaviour, and then asked them to state how well their answers reflected their ideal preferences and, if not, why.
To structure these findings, we classify participants according to their general consent answers: always accept, mostly accept (>= 75\%), mixed consent, mostly reject (>= 75\%), always reject.


When asked what they based their choices on, the answers touched on eleven different topics.
The four `always accept' participants cited a general apathy towards privacy concerns and ``\textit{just did it to make the window go away}''.
The one participant that `always rejected', no matter whether that required more effort, argued that they would only accept data collection if it was to use a particular feature offered by the site.
The eleven participants categorised as `mostly reject' heavily emphasised a disagreement with the practice of tracking in general and stated they would only consent to have their data collected if it was for websites they trusted. Two of those also mentioned that they did not feel a need for any personalisation. 
The participants that fell into the `mostly accept' and `mixed consent' category were more diverse. 
Most often mentioned were pragmatic reasons such as just wanting to get to the site as quickly as possible, not believing the controls were meaningful, and not wanting to lose any functionality.
Eight decided based on trust, whether it was the website or the vendors, and the sensitivity of the data they would be submitting (e.g., banking information). 
One participant stated that they relied on other methods to protect their privacy, so did not care that much about their pop-up answers: ``\textit{I tend to vary my devices/browsers/accounts/use incognito and duckduckgo a lot, I'm not too worried about my data being tracked to every detail}.''

After being shown a visualisation of their actual consent behaviour and asked if it matched their ideal settings, the responses were predominately that it did not.
Only those falling into the two extreme categories -- `always accept' and `always reject' -- all indicated they agreed (3) or strongly agreed (2) with their answers.
For the remaining three categories, the sentiments were mostly spread evenly along the spectrum, with 11 somewhat agreeing, 3 neither agreeing nor disagreeing, 7 somewhat disagreeing, 1 disagreeing, and 3 strongly disagreeing.

The 25 participants who indicated their behaviour did not match their ideal privacy settings were asked to explain what the reason for this difference was.
Participants mentioned desires such as just wanting more privacy (``\textit{I would rather companies not collect any information}''); the fear of unknown consequences of opting-out (``\textit{I didn't want to risk the website not working after that}''); and not knowing what their ideal preferences even are.
The most common reason mentioned, however, was the interface design.
Participants lamented the fact that pop-ups stand in the way of their primary goal (accessing a service), that the frequency of the pop-ups caused frustration and consent fatigue, and even the perception that the pop-up ``\textit{forced them to accept}'' -- even though these options were available on the second page.

\subsection{Interim Discussion}
The experimental results indicate how two of the most common consent interface designs -- not showing a `reject all' button on the first page; and showing bulk options before showing granular control -- make it more likely for users to provide consent, violating the principle of ``freely given''\footnote{It should be noted this data alone is not enough to establish legal compliance.}.
The notification style, on the other hand, appears to have no effect on the answer, but possibly a large effect on whether an answer is given at all, suggesting that a non-blocking mechanism provides a desired third consent option to users: a neutral middle-ground.
The qualitative reflections of the participants, however, put into question the entire notice-and-consent model not because of specific design decisions but merely because an action is required before the user can accomplish their main task and because they appear too frequently if they are shown on a website-by-website basis.


\subsection{Limitations}
The participant sample is by no means representative of the general population in the United States: they are almost all young and university-educated, and recruited primarily through an emailing list of a computer science department. 
Arguably, this means that our results describe a ``best case scenario'': these participants should be more knowledgeable about privacy issues and better equipped to understand consent interfaces than the average web user.

There are a number of confounding variables that could have affected the participants' answers.
First, although the condition order was counterbalanced, we cannot guarantee that the participants were actually exposed to them in that order (e.g., if they opened multiple tabs in a row and visited them anachronistically), meaning order effects might not be controlled for.
Second, because we showed the same pop-up to each participant until we recorded four answers per interface, some participants were exposed to the different conditions more often than others.
Lastly, participants might have also encountered ``real'' pop-ups at the same time as the injected ones if the website they were visiting was within the territorial scope of the GDPR.

While the GDPR is a European policy, our experiments were conducted in the United States. 
These populations have been exposed to different legal regimes and different consent controls over the year, something which we expect has affected their mental model of these kind of pop-ups and accordingly, how they answer them.
This might influence the extent to which these findings can be generalised to a European population, and thus how they should be used to inform EU policy changes.
\section{Discussion and Conclusion}

The results of our empirical survey of CMPs today illustrates the extent to which illegal practices prevail, with vendors of CMPs turning a blind eye to --- or worse, incentivising ---- clearly illegal configurations of their systems. Enforcement in this area is sorely lacking. Data protection authorities should make use of automated tools like the one we have designed to expedite discovery and enforcement. Designers might help here to design tools for regulators, rather than just for users or for websites. Regulators should also work further upstream and consider placing requirements on the vendors of CMPs to only allow compliant designs to be placed on the market. Such enforcement may be possible as the Court of Justice indicates that plugin system designers can be `joint controllers' along with websites~\cite{fashionid,mahieuResponsibilityDataProtection2019a,vanalsenoy}, and the UK's ICO indicates it may be willing to force advertising trade bodies to alter their standards~\cite{informationcommissionersofficeUpdateReportAdtech2019}. If this is the case, regulators must carefully consider how to build a robust and well-maintained evidence base for user-centric CMP design.

A core takeaway from the user study is that placing controls or information below the first layer renders it effectively ignored. This leaves a few options for genuine control of tracking online. If the notice-and-consent model is to continue, it may be necessary to declare that, for example, consent can never be valid with the presence of the (on average) hundreds of third parties we have shown  data is sent to and cookies laid by today. This would mean consent would only be valid if a compact but representative and rich description can be placed on the first layer, and could certainly be a possible direction for the Court of Justice to consider if they interpret the principles of data protection in a future case. 

An alternative approach would be to overhaul the design pattern of the consent banner or barrier, and have richer, more durable ways to set preferences, potentially within the browser. The key is that such browser settings would be legally binding, rather than weak and self-regulatory in nature. Yet the current heavy lobbying around the EU's draft ePrivacy Regulation has centred in part on adtech firms trying to prevent browser settings having legally binding effect --- part of an ongoing drama for many years about the potential legal status of `Do Not Track' signals~\cite{kamaraNotTrackInitiatives2016a}. Designers have a role here: how can users reflect on tracking across the Web, rather than on a per-site basis? If users are not to automatically reject everything, how can advertisers negotiate and present them with reasons that they should consent? Might there be a role for delegation of preferences to a trusted civil society actor, and what kind of relationship, information and interaction might the user have with these? We invite and encourage researchers to bring their skills and views to bear on these important, current issues at the confluence of regulation, design and fundamental rights.

% ...

\section{Acknowledgments}
Our thanks to Jos\'e Juan Dominguez Veiga and Kristian Borup Antonsen for their invaluable assistance programming the scraper and extension, and to Luke Taylor and Benjamin Cowan for their advice with the statistical analysis.

Midas Nouwens was supported by the Aarhus Universitets Forskningsfond and the Danish Agency for Science and Higher Education.
Ilaria Liccardi  was supported by the William and Flora Hewlett Foundation and the NSF 1639994: Transparency Bridges grant.
Michael Veale was supported by the Alan Turing Institute under EPSRC grant no. EP/N510129/1.

\balance{}

\bibliographystyle{SIGCHI-Reference-Format}
\bibliography{bibliography.bib}

\end{document}